\documentclass[a4paper]{article}

\usepackage{18lomcon}        
\usepackage{cite}             
\usepackage{epsfig}           
\usepackage{epstopdf}

\begin{document}


\title{LHC HUNTING THE ODDERON: IS IT REALLY CATCHED?\footnote{The talk at the 19th Lomonosov Conference on Elementary Particle Physics, August 22-28 2019, Moscow. To be published in the Proceedings of the Conference.} }

\author{Vladimir A. Petrov \email{Vladimir.Petrov@ihep.ru}}

\affiliation{A. A. Logunov Institute for High Energy Physics, NRC "Kurchatov Institute", 152281, Protvino, RF}


\date{}
\maketitle


\begin{abstract}
   We give a short survey of recent LHC experiments and related theory topics concerning the Odderon problem.
\end{abstract}

\section{Introduction}
 The Odderon concept  was pushed forward  in 1973 \cite{Nic} and  embodied a hypothesis that the leading C-even agent dominating high energy behaviour of the strong interaction cross-sections, the famous Pomeron, may have a C-odd counterpart which can give non-negligible contributions at high energies in contrast  to secondary Reggeons $ \rho,\omega $ etc rapidly  dying off with the energy growth.
One of the motivations was an extension of the old Chew-Frautschi imperative  of "maximum strength" for strong interactions at high energies \cite{Che}  from the asymptotic constancy of the total cross-sections to the functional saturation of the Froissart bound implying their $ \sim \ln ^{2} s $ behaviour but, more than that, the similar saturation of the upper bound for the differences of "C-conjugated " processes (e.g. for $ \Delta\sigma = \sigma_{tot}^{\bar{p}p} - \sigma_{tot}^{pp}$) which is bounded above ( \textit{modulo modulus}) by $ \sim ln s $.
Thus, the "maximal Odderon" hypothesis implies the violation of the Pomeranchuk theorem (in the sense of differences)which asks for  $ \Delta\sigma\rightarrow 0 $ and, in a more general context, the Gribov "principle of asymptotic universality" \cite{Gri} extended to the case of indefinitely rising cross-sections.

From a (maybe na\"{i}ve) physical viewpoint the idea of the maximal Odderon implies, in particular, that the difference between the proton-proton and anti proton-proton interactions gets larger and larger when the average distance (impact parameter $ b $) between the colliding particles ("interaction range") increases\footnote{We mean the well-known increase of the forward slope $ B(s)\approx \langle b^{2} \rangle/2 $ with the energy growth.}.

In very simple words, colliding particles are believed to discern the difference in their inner structure the better, the farther they are from each other.
 
Nonetheless, such an unorthodox option  \cite{Nic}, albeit not very appealing from the underlying physics viewpoint, seems to be fairly conceivable in a formal sense and even shows up a certain elegant symmetry for forward amplitudes
because it implies that "at high energies"
\begin{equation}
T^{+}(s,0) = iA\cdot ln^{2}(s\cdot exp(-i\pi/2)),
\end{equation}
\begin{equation}
T^{-}(s,0) = B\cdot ln^{2}(s\cdot exp(-i\pi/2))
\end{equation}
where 
\[T^{\pm }(s,0) = \frac{1}{2}[T^{\bar{p}p}(s,0) \pm T^{pp}(s,0)] .\]

Let us note that not every pair of "C-conjugated" processes  can be associated with the Odderon exchange\footnote{The term "exchange" is used in a wide sense meaning only the general quantum number and energy-momentum exchange irrelevant to a concrete mechanism.}. 
E.g., the pair of processes
$ \pi^{\pm} p \rightarrow \pi^{\pm} p $ is a counter example. We, however, will deal mostly with 
(anti)proton-proton scattering which is under active discussion in relation with a rich experimental material accumulated by now by the collaborations TOTEM and, partly by ALFA(ATLAS), for $ pp $ collision (LHC) and by the D0 Collaboration (Tevatron) for $ \bar{p}p $ collisions.

\section{Experimental searches}
\subsection{Forward observables}

"Forward observables" mean differential cross sections $ d\sigma/dt $ for small momentum transfers $ t\rightarrow 0 $, total cross sections (since they are proportional to the imaginary part of the forward scattering amplitude $ T(s,0) $) and the ratio $ ReT(s,0)/ImT(s,0)\doteq \rho (s) $.

The very beginning of introducing the maximal Odderon idea  was marked by expectations of a dramatic, both qualitative and quantitative, change of mutual relation between the total $\bar{p}p  $ and $ pp $ cross-sections, viz., the cross-over 
changing $\bar{p}p $ dominance to the $ pp $ one. With time the predicted energy where such an event should occur moved from $ \mathcal{O}(20) GeV $ to the recent estimate near $ 300  $ GeV.
In the absence of "simultaneous" measurements of  $\bar{p}p $ and  $ pp $ in the post-ISR energy region it is quite difficult to argue about such a cross-over: the difference of order $ \mathcal{O}(1 mb) $ is expected at c.m.s. energies no less than $ 100 $ TeV, far beyond any realistic plans. 

Other Odderon options could give some evidence if the intercepts of corresponding j-plane singularities
while being lower than that of the Pomeron would lie higher than the secondary (quarkic) Regge trajectories. This could, in principle, give the slower decrease of $ \Delta\sigma_{tot}=\sigma^{\bar{p}p}_{tot} - \sigma^{pp}_{tot}$ than expected from the secondary trajectories. However, again the lack of the $ \bar{p} p $ data in the TeV region does not allow such a test of the Odderon existence.

Nonetheless, there are other features of the maximal Odderon which are believed to be tentatively caught with now existing means. For instance, a latest embodiment of the maximal Odderon doctrine\cite{NM} , the "Froissaron- Maximal Odderon" model  (FMO), yielded the value of the parameter $ \rho $ almost exactly coinciding with its value published by the TOTEM Collaboration \cite{Ant}. 
It was a reason to claim that the TOTEM Collaboration managed to finally catch the hitherto elusive Odderon.

 Leaving aside some serious theoretical flaws in the  mentioned above "maximal Odderon" model (to be discussed elsewhere) it should be noted that there appeared publications  (see, e.g.\cite{Sel}) successfully describing the same TOTEM data without any C-odd forces comparable with the Pomeron strength. 

Moreover, the very experimental values of the signal quantities, $ \rho $ and $ \sigma_{tot} $, presented in \cite{Ant} were questioned. The matter is that the procedure of extraction of these parameters from the data depends essentially on the way of accounting for Coulomb-nuclear interference (CNI) which was considered differently by different authors.

For instance, it was demonstrated explicitly in Ref.\cite{EPT} that the use of a modified CNI formula \cite{Pet} leads to higher values of $ \rho $ quite compatible with early predictions.

 Howbeit, there are old arguments that small-angle observables are not appropriate to search for the Odderon effects \cite{PS} and even maximally high Odderon intercept admissible by unitarity can lead, at best, to an asymptotically constant limit of $ \Delta\sigma_{tot}=\sigma^{\bar{p}p}_{tot} - \sigma^{pp}_{tot}$ \cite{Fi}.
 
 We have also to add that the C-odd exchange of gluonic origin is under active theoretical studies both in perturbative and nonperturbative (lattice) QCD  quite for a long time (see, e.g. the review in \cite{Ew})  but, unfortunately,  no definite conclusions were obtained by now, as seen from the recent publication \cite{Bar}.

\subsection{Non forward observations}

The study of non forward observables ( mostly $ d\sigma/dt $ at $ t $ beyond the forward peak region) in order to detect Odderon effects also needs, like in the forward case, the data both in $ pp $ and $\bar{p}p$ channels.
This need has been accomplished at the ISR where a comparison between $ d\sigma^{pp}/dt $
and $ d\sigma^{\bar{p}p}/dt $ at $ \sqrt{s} = 53 $ GeV  was made \cite{ISR} . Both cross-sections coincided with a very good accuracy over the entire $ t $ range except the vicinity of the dip of $ d\sigma^{pp}/dt $ (shoulder of $ d\sigma^{\bar{p}p}/dt $) where

\[d\sigma^{\bar{p}p}/d\sigma^{pp} \approx 4.5 \pm 1.5.\]

In the TeV energy domain we have no $ d\sigma/dt $ for $ \bar{p}p $ and $ pp $ measured at the same energy.We have, however, such a pair measured at energies $ 1.96 $ TeV($\bar{p}p $)\cite{Ab}  and $ 2.76 $ TeV ($ pp $)\cite{TO2}. Taking into account a certain "slowness" of the energy dependence of the diffractive scattering ($ \sigma_{tot}^{pp} $ increases only 2.75 times while the energy increases $ \sim $ 260 times) it seems reasonable to compare $ d\sigma/dt $ for $ \bar{p}p $ and $ pp $ at energies indicated above  and which differ only 1.4 times. 

We have for the ratio in question
\[d\sigma^{\bar{p}p}/d\sigma^{pp} \approx 1.2 \pm 0.3.\]

Thus we see that  $d\sigma^{\bar{p}p}/dt $ still prevails over  $ d\sigma^{pp}/dt $ at  TeV energies but this prevalence has decreased almost four times.

Whether this C-odd effect can be described with use of the secondary quarkic trajectories only or it asks for a one or another Odderon option depends on the model and at the time being there is no consensus on this subject.

 \section{New prospects}
 
 Diffraction processes are not limited to elastic scattering.
 For the subject in discussion, the Odderon, very interesting prospects promise
 so called "central diffractive processes" in which both target and projectile remain intact after the collision (sometimes their diffractive excitation is admitted) but with some fraction of their energies invested
into production of some state with small rapidities in the c.m.s.
For example, we can study central production processes
\[p + p\rightarrow p \oplus C_{+} \oplus p \]
or 
\[A + p \rightarrow A \oplus C_{+} \oplus p\]
( $ \oplus $ means a large (pseudo)rapidity gap, $ A $ stands for a heavy ion and $ C_{+} $ for a C-even state) in which centrally produced C-even state (e.g.$ f_{2} $ \cite{Go}) my be produced in a subprocess where weakly virtual gamma quantum (enhanced in case of nucleus $ Z^{2} $ times ) interact with the Odderon \cite{Kh}.
Another opportunity is the similar central production of a C-odd state ( e.g. $ \phi, \omega, J/\psi $) where the driving subprocess can be the Pomeron-Odderon collision.
Recently started project CT-PPS \cite{CT} inspires some optimism in this regard.
\section*{Conclusions}
 So we have to admit that in spite of a rich experimental material and theoretical efforts the Odderon still remains elusive. No doubts, C-odd partners of the Pomeron(s) should exist - it is hard to imagine that something could prohibit its existence in QCD.
 
 The Odderon problem (along with the Pomeron problem) still stays, as it is seen, a quite a difficult task on the way to understanding the laws of strong interactions at high energies.

\section*{Acknowledgments}

I am very grateful to Professor A. I. Studenikin and his dynamic young team for the excellently organized conference.

\end{document}